\documentclass[aps,prb,twocolumn,superscriptaddress,showpacs,floatfix,longbibliography]{revtex4-1}
\usepackage{amsmath,amssymb,amsfonts,float,graphics,epsfig,epstopdf,color,verbatim,tabularx,bm,multirow,appendix}
\usepackage[utf8]{inputenc}
\usepackage[T1]{fontenc}
\usepackage{xcolor}
\usepackage{verbatim}

\IfFileExists{newtxtext.sty}
{\usepackage{newtxtext,newtxmath}}
{\IfFileExists{stix.sty}
	{\usepackage{stix}}
	{\IfFileExists{mathptmx.sty}
		{\usepackage{mathptmx}}{} } }

\usepackage{textcomp}
\usepackage{bm}

\IfFileExists{siunitx.sty}{\usepackage{booktabs,siunitx}}{}

\usepackage{color}
\definecolor{LinkColor}{rgb}{0.256,0.439,0.588}
\usepackage{hyperref}
\hypersetup{
	pdfauthor={Yuzhi Liu},
	pdftitle={paper},
	colorlinks=true,
	citecolor=LinkColor,
	linkcolor=LinkColor,
	urlcolor=LinkColor
}

\usepackage{pifont}

\providecommand{\U}[1]{\protect \rule{.1in}{.1in}}

\begin{document}
\title{Fermion enhanced first-order phase transition and chiral Gross-Neveu tricritical point}
\author{Yuzhi Liu}
\affiliation{Beijing National Laboratory for Condensed Matter Physics and Institute of Physics,Chinese Academy of Sciences, Beijing 100190, China}
\affiliation{School of Physical Sciences, University of Chinese Academy of Sciences, Beijing 100190, China}
\author{Zi Yang Meng}
\affiliation{Department of Physics and HKU-UCAS Joint Institute of Theoretical and Computational Physics, The University of Hong Kong, Pokfulam Road, Hong Kong, China}
\affiliation{Beijing National Laboratory for Condensed Matter Physics and Institute of Physics,Chinese Academy of Sciences, Beijing 100190, China}
\author{Shuai Yin}
\affiliation{School of physics, Sun Yat-Sen University, Guangzhou, 510275, China}

\begin{abstract}
The fluctuations of massless Dirac fermion can not only turn a first-order bosonic phase transition (in the Landau sense) to a quantum critical point, but also work reversely to enhance the first-order transition itself, depending on the implementation of finite size effects in the coupling corrections. Here, we report a case study of the latter by employing quantum Monte Carlo simulation upon a lattice model in which the bosonic part featuring the Landau-Devonshire first-order phase transition and Yukawa coupled to the Dirac fermions. We find that the parameter range for the first-order phase transition becomes larger as the Yukawa coupling increases and the microscopic mechanism of this phenomena is revealed, at a quantitative level, as the interplay between the critical fluctuations and the finite-size effects. Moreover, the scaling behavior at the separation point between the first-order and the continuous phase transitions is found to belong to the chiral tricritical Gross-Neveu universality. Our results demonstrate that the interplay of massless Dirac fermions, critical fluctuations and the finite size effects could trigger a plethora of interesting phenomena and therefore great care is called for when making generalizations.
\end{abstract}
\date{\today}

\maketitle

\section{Introduction}
\label{sec:i}
Fluctuations play vital roles in both first-order and continuous phase transitions~\cite{Landau,Fisher1967,Wen,Sachdev}. It was realized that the self-similar fluctuating modes are responsible for the scaling behaviors in the second-order phase transition by Wilson's renormalization group theory~\cite{Wilson1974}, which consequently brought in the notion of the universality class -- one of the organization principles in statistical and condensed matter physics. Moreover, fluctuations can even change the order of the phase transitions. For example, the Coleman-Weinberg mechanism showed that the fluctuation of gauge field can turn a continuous phase transition into a first-order one~\cite{Coleman1973,Ihrig2019}. On the contrary, the theory of the deconfined quantum critical point proposed that the fluctuations from the fractionized spinons and emergent gauge field can round a first-order transition (in the Landau sense) between two ordered phase into a continuous one~\cite{Sandvik2007,Nogueira2007,Melko2008,Block2013,Lou2009,Pujari2013,Nahum2015A,Shao2016,Nahum2015B,YQQin2017,Sreejith2019,NSMa2018,NSMa2019}.
Landau-Ginzburg(LG) model is the typical model to achieve a continuous phase transition. Landau-de Gennes~\cite{Toledano} model introduce the cubic term into LG and Landau-Devonshire model~\cite{Toledano,Devonshire,Blume,Capel} increase the $\phi^6$ term based on LG with minus $\phi^4$ term. Both of them are effective model for the first-order transiton. In a similar spirit with deconfined quantum critical point, fluctuations of Dirac fermions can also soften the Landau-de Gennes and the Landau-Devonshire first-order transition in the bosonic sector into continuous ones, which are dubbed as the type-I~\cite{TCLang2013,Li2017,Scherer2016,Meng2016,Classen2017,Jian2017A,Sato2017,Torres2018} and type-II fermion-induced quantum critical points (FIQCP)~\cite{Yin2020}.

The aforementioned model studies~\cite{Sandvik2007,Nogueira2007,Melko2008,Block2013,Lou2009,Pujari2013,Nahum2015A,Shao2016,Nahum2015B,YQQin2017,Sreejith2019,NSMa2018,NSMa2019,TCLang2013,Li2017,Scherer2016,Meng2016,Classen2017,Jian2017A,Sato2017,Torres2018} are usually carried out numerically on finite lattice sizes. It is well-known that the finite size (using $L$ to denote the linear lattice size) provides a natural infrared truncation in the long wavelength fluctuation, and one shall perform finite-size scaling (FSS)~\cite{Cardy} to extract the critical properties of the universality, i.e., treating $L$ as a tunable relevant scaling variable to estimate the critical point and exponents. In particular, it was shown that the scaling form of the FSS should be drastically amended near the deconfined quantum critical point as a result of the appearance of the dangerously irrelevant scaling variable~\cite{Shao2016}. On the other hand, a controlled FSS analysis in the Dirac fermion induced quantum critical point is still rare~\cite{Liu2019}, and it is our first motivation in this work to address this issue.

The critical properties of the interacting Dirac fermion systems have attracted attentions from condensed matter to high-energy physics communities, not only in the discussion of quantum electrodynamics with fermionic matter~\cite{XYXu2019,Karthik2019,Dupuis2019,WeiWang2019,Zerf2020,Janssen2020}, but also in that the Dirac fermion drives the Wilson-Fisher fixed point into the chiral Gross-Neveu fixed point~\cite{TCLang2013,Mihaila2017, Zerf2017,Gracey2018,Gracey2018A}. Although enormous investigations have been devoted to this issue~\cite{Torres2020,Hennadii, Herbut2006,Honerkamp2008,Herbut2009,Strack2010,Yao2015,Ihrig2018,Mihaila2017,Zerf2017,YYHe2018,Meng2019,Lang2019,herbutlor,Schuler2019,Liu2019,Gross1974,Rosenstein1993,JanssenHerbut2014,Gracey2018,Knorr2018}, the numerical verification for type-II FIQCP is still lacking.
Remarkably, recent studies based on the field-theoretical effective model propose that type-II FIQCP can also feature new tricritical behaviors, controlled by the chiral tricritical point (CTP)~\cite{Yin2018}. This CTP separates the conventional Landau-Devonshire first-order transition from the type-II FIQCP~\cite{Yin2020}, and the universal scaling behavior near this CTP is quite different from its pure bosonic counterpart. A numerical verification on such CTP constitutes our second motivation.

In this paper, we hit the two birds with one stone by investigating numerically a lattice model, which consists of a spin (boson) part hosting the Landau-Devonshire first-order transition, a massless Dirac fermion part and the coupling between them. The numerical approach is based on the determinant quantum Monte Carlo (DQMC) method~\cite{BSS1981,Assaad2008,Mengreview2019,Liu2019,Torres2020}. Although the theory of the type-II FIQCP predicts that the fermion fluctuation can soften the boson first-order transition into a continuous one~\cite{Yin2020}, here we find apparently things can also go to the opposite direction in that the range of the first-order transition is extended due to the coupling with Dirac fermions. To explain this observation, we develop a modified mean-field theory to study the effective coupling in the free energy and reveal that this anomalous phenomenon is induced by the interplay between critical fluctuations and the finite-size effects, in a quantitative manner. Moreover, we pinpoint the tricritical point separating the first-order and the continuous phase transitions, and numerically verify that this CTP acquires the critical exponents of chiral Gross-Neveu universality, confirming the renormalization group predictions~\cite{Yin2018}.

The rest of the paper is organized as follows. Sec.~\ref{sec:ii} introduces the lattice model and DQMC methodology. The numerical results are shown in Sec.~\ref{sec:iiiA} where we demonstrate the Dirac fermion enhanced first order transition. To explain it, in Sec.~\ref{sec:iiiB} a modified mean-field analysis is presented. In Sec.~\ref{sec:IV} the position and critical exponents at the Gross-Neveu CTP are revealed with FSS upon numerical data. Finally, a summary is given in Sec.~\ref{sec:V}.

\section{lattice model and numerical method}
\label{sec:ii}
\begin{figure}[htp!]
\includegraphics[width=\columnwidth]{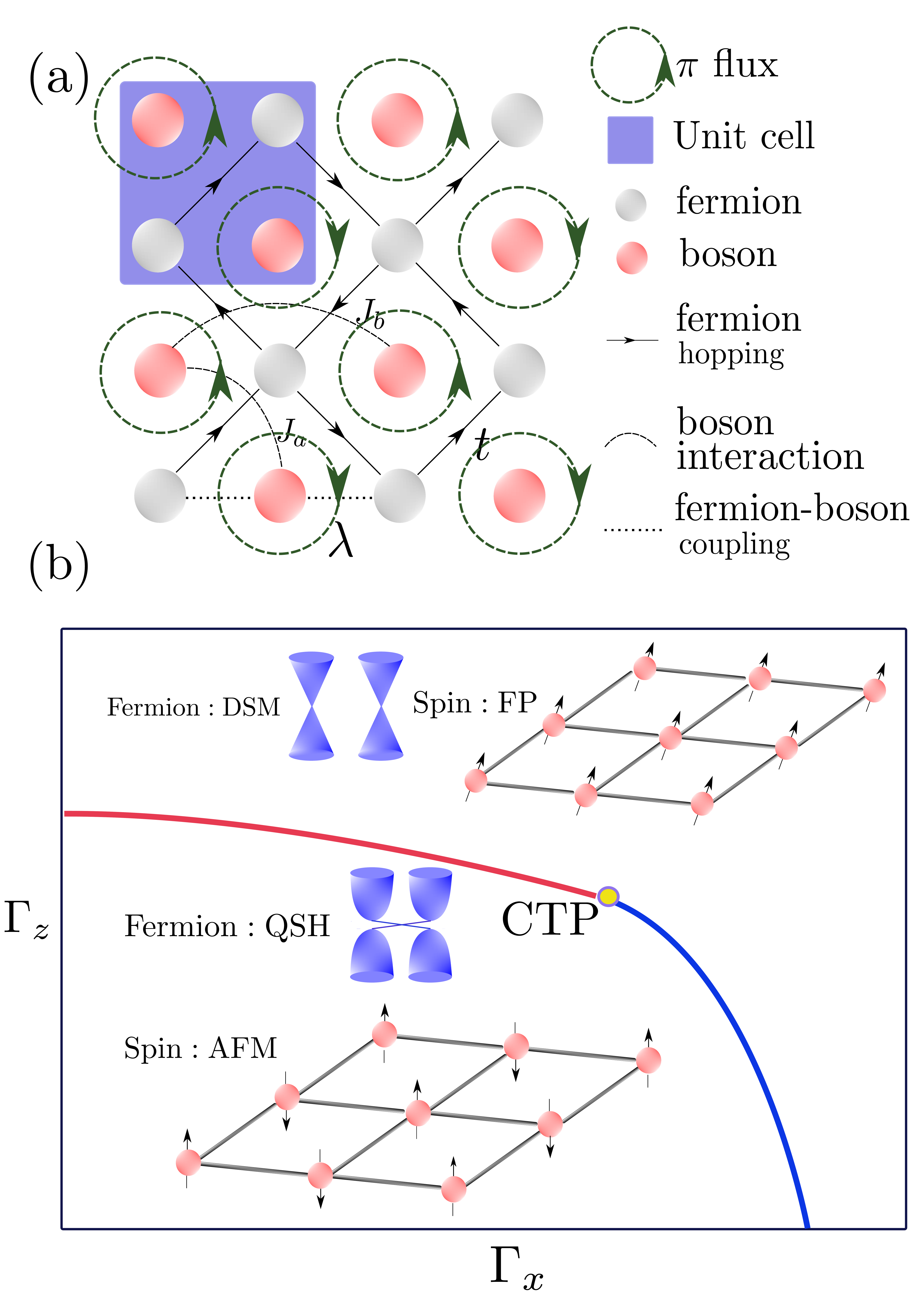}
\caption{The lattice model and the ground-state phase diagram. (a) The gray lattice sites and red lattice sites respectively present the fermion and boson sites. One unit cell therefore contains two fermion sites and two boson sites. The solid line with arrow means the fermion hopping. The dash straight line means fermion-boson coupling as Eq.~\ref{eq:eq2}. The dash curve between the bosonic sites is boson interaction and the dash circle with arraw on each fermionic plaquette means the $\pi$-flux. (b) The schematic ground state phase diagram, AFM and FP phases of the Ising spins are separated by a continuous (blue line) or first-order (red line) phase transitions where they meet at the tricritical point (CTP) of the model. In the AFM and FP phases, fermion is inside the quantum Spin Hall (QSH) and Dirac semimetal (DSM) states due to the coupling with the bosons.}
	\label{fig:fig1}
\end{figure}

The lattice model is comprised of Dirac fermions, Ising spins (bosons) and their coupling, on the square lattice. As shown in Fig.~\ref{fig:fig1}(a), the bosonic part reads~\cite{Kato2015}
\begin{eqnarray}
\label{eq:eq1}
\begin{split}
H_{\text{Boson}} &=& J_{a}\sum_{\langle p,q \rangle}\sigma^{z}_{p}\sigma^{z}_{q} - J_{b}\sum_{\langle \langle p,q \rangle \rangle}\sigma^{z}_{p}\sigma^{z}_{q}\\
 &&-\Gamma_{z}\sum_{p}\sigma^{z}_{p} - \Gamma_{x}\sum_{p}\sigma^{x}_{p},
\end{split}
\end{eqnarray}
in which the Pauli matrices $\sigma_{p}^{z/x}$ represent a local spin at the bosonic site $p$, $J_a$ represents the nearest antiferromagnetic (AFM) interaction, $J_b$ represents the next-nearest ferromagnetic (FM) interaction, and $\Gamma_{x}$ is the transverse field, $\Gamma_{z}$ is the longitudinal field.

The fermion part reads~\cite{Harris1989,YYHe2018,Liu2019}
\begin{equation}
\label{eq:eq2}
H_{\text{Fermion}}=\sum_{\langle i, j \rangle,\sigma_{f}}-t_{ij}\text{e}^{i\sigma_{f}\theta_{ij}}c_{i,\sigma_{f}}^{\dagger}c_{j,\sigma_{f}} + \mu\sum_{i}n_{i} +h.c.,
\end{equation}
in which $c_{i,\sigma_{f}}$ ($c^{\dagger}_{i,\sigma_{f}}$) is the fermionic annihilation (creation) operator at the fermionic site $i$ with spin $\sigma_{f}=\pm 1/2$, and the phase $\theta_{ij}$ is set to be $\theta_{ij}=\pi/4$ which allows a $\pi$-magnetic-flux on each fermionic plaquette and supports two Dirac points in its energy bands~\cite{Lieb1994}. $n_i$ is density of fermion and $\mu$ is chemical potential. We set $\mu=0$ for half filling of fermions.

The coupling between Eqs.~(\ref{eq:eq1}) and (\ref{eq:eq2}) is
\begin{equation}
\label{eq:eq3}
H_{\text{Coupling}}=\sum_{\langle \langle i,j \rangle \rangle,\sigma_{f}}\lambda_{ij}\sigma^{z}_{p}c^{\dagger}_{i,\sigma_{f}}c_{j,\sigma_{f}} + h.c.,
\end{equation}
in which $\lambda_{ij}$ represents the coupling strength. Thus the total Hamiltonian is
\begin{equation}
\label{eq:eq4}
H=H_{\text{Boson}}+H_{\text{Fermion}}+H_{\text{Coupling}}.
\end{equation}
throughout the paper, we set $t_{ij}=t=1$ as the energy unit and $\lambda_{ij}=\lambda$ the same on every bond.

The schematic phase diagram of the model, spanned by the axes of $\Gamma_x$ and $\Gamma_z$, is shown in Fig.~\ref{fig:fig1} (b). In the absence of the coupling to the Dirac fermions, i.e., $\lambda=0$, the pure spin model with fixed $J_a$ and $J_b$ has two phases~\cite{Kato2015}
. For small $\Gamma_x$ and $\Gamma_z$, the system is in a AFM phase, while for large $\Gamma_x$ or $\Gamma_z$, the system is in a fully-polarized (FP) phase. By tuning $\Gamma_x$ and $\Gamma_z$, there is a phase transition between these two phases. For small $\Gamma_z$, the phase transition is continuous and belongs to the $(2+1)$D Ising universality class as denoted by the blue line. For large $\Gamma_z$, the phase transition is first-order as denoted by the red line. It was shown that when $J_a=J_b$, this first-order transition can be casted into the Landau-Devonshire effective model with a negative quartic coupling~\cite{Devonshire, Kato2015}. In addition, there is a quantum tricritical point (CTP) separating the first-order and the continuous phase transition. In the following, we also perform the simulation at $J_a=J_b$, since in this case the uniform part of $\langle\sigma_z\rangle$ can be treated as a background field rather than a dynamical field and $H_{\text{Boson}}$ has been solved with quantum Monte Carlo simulation in Ref.~\onlinecite{Kato2015}.

$H_{\text{Fermion}}$ is the $\pi$-flux model which gives rise to two Dirac cones at $(\pi, 0)$ and $(0, \pi)$ in the Brillouin zone (BZ). As shown in Fig.~\ref{fig:fig1} (a), we couple a pair of next-nearest-neighbor fermion sites in $H_{\text{Coupling}}$ which the sign relys on the spin at the bosonic site. In the AFM phase of Ising spins, a mass term can be generated for fermions which gap out the Dirac points, transforming the Dirac semimetal (DSM) into a dynamically-generated quantum spin Hall insulator (QSH)~\cite{YYHe2018}. In the FP phase, bosonic field still keep the Dirac cone at $(\pi, 0)$ and $(0, \pi)$ intact but renomalized the high energy parts of the bands in the BZ. Meanwhile, the original 3D Ising Universility class between AFM phase and FP phase is replaced by chiral Ising Gross-Neveu universality class.This result has been numerical revealed by some of the present authors in Refs.~\onlinecite{YYHe2018,Liu2019}.

In the presence of the coupling to the Dirac fermions, there are two main theoretical predictions: one is that the fermion fluctuation can drive the first-order transition into a continuous one~\cite{Yin2020}. This may indicate that region of the first-order transition should shrink as long as the coupling to the Dirac fermion is introduced. Surprisingly, in the following, we will show a contrary phenomenon has occurred in the actually DQMC simulation and provide an explanation. The other is that the universality class of the tricritical point is drastically changed by the gapless Dirac fermions~\cite{Yin2018}. The first-order and the continuous transition belonging to the chiral Ising  Gross-Neveu univerality class is seperated by this CTP which is in the chiral tricritical Ising unversality class. 


The compuation of $H$ can be carried out without sign problem in DQMC method, and we present the detailed implementation in Appendix.~\ref{app:appA}.

\section{Fermion-enhanced first-order transition}
\label{sec:iii}
In Sec.~\ref{sec:iiiA}, we first present numerical results showing the range of the first-order phase transition in our model is actually extended rather than shrunk, i.e. the first order transition line (the red line in our Fig.~\ref{fig:fig1} (b)) in the $\Gamma_x-\Gamma_z$ phase diagram extends a bit towards larger values of $\Gamma_x$ and $\Gamma_z$ compared with that of the bare spin model~\cite{Kato2015}, seemingly contrary to the expectation. Then in Sec.~\ref{sec:iiiB} we will give an self-consistent explanation based on a modified mean-field analysis to reconcile the puzzle.

\subsection{Numerical results}
\label{sec:iiiA}
In the DQMC simulation, we choose the parameter $J_a=J_b=1$ and explore the phase transition properties by scanning the $\Gamma_z$ for different $\Gamma_x$.

\begin{figure}[htp!]
\includegraphics[width=0.9\columnwidth]{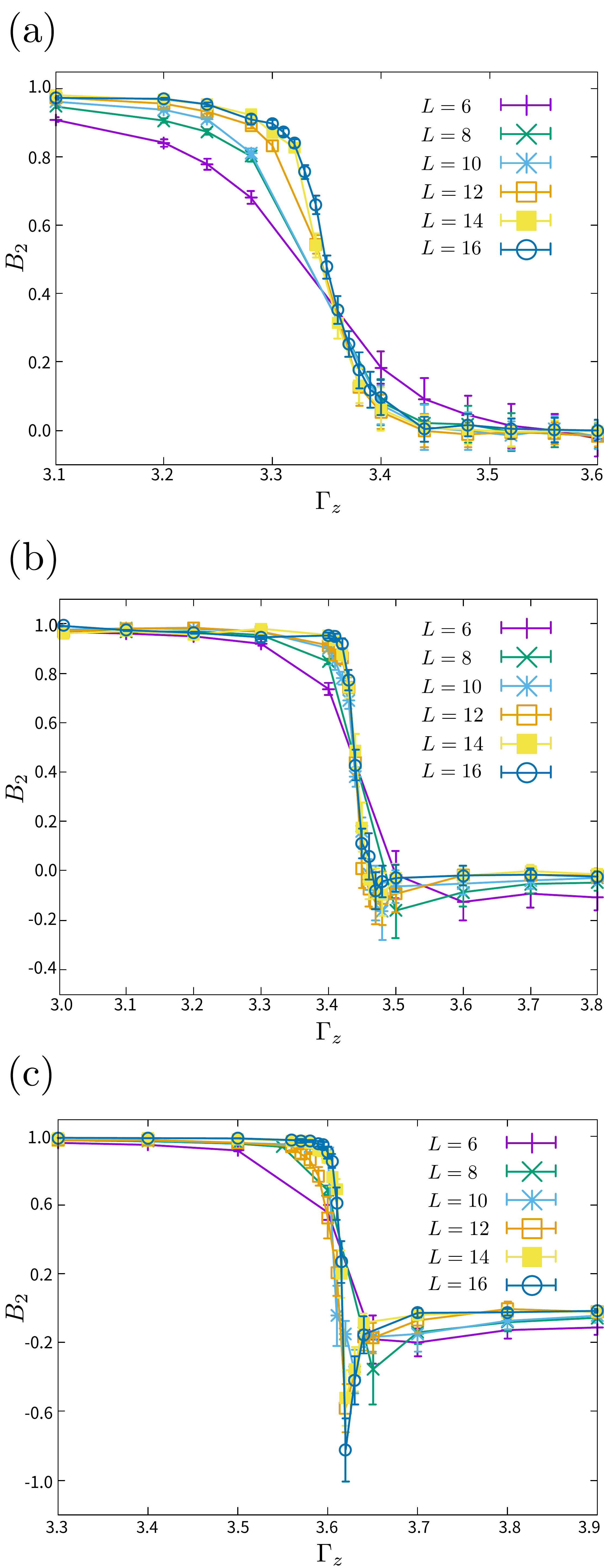}
\caption{Binder ratio  $B_{2}$ in different couplings $\lambda$. (a) Binder ratio of continue phase transition for the case of $\lambda=0.0$, $\Gamma_x=4.4$. (b) Binder ratio close to the tricritical point for the case of $\lambda=0.0$, $\Gamma_x=4.0$. (c) Binder ratio of first order transition for the case of $\lambda=0.3$, $\Gamma_x=4.0$.}
	\label{fig:fig2}

\end{figure}

For finite-size systems, phase transition properties can be reflected by the behavior of the Binder ratio near the phase transition point. In the present case the Binder ratio is given as
\begin{equation}\label{eq:eq5}
B_2 = \frac{3}{2}\left(1-\frac{1}{3}\frac{\langle m^4 \rangle}{\langle m^2 \rangle^2}\right),
\end{equation}
in which
\begin{equation}\label{eq:eq6}
m^{2}=\frac{1}{L^2}\sum_{k,l}(-1)^{\alpha_1-\alpha_2}\sigma^{z}_{k,\alpha_1}\sigma^{z}_{l,\alpha_2}
\end{equation}
with $L$ being the lattice size,$k, l$ being the unit cell in which $\alpha_1, \alpha_2$ is the site of the unit cell, such that the staggered magnetization of the AFM phase is measured. In the ordered phase $B_2 \rightarrow 1$, while in the disordered phase $B_2 \rightarrow 0$. In continuous phase transitions, if the scaling correction can be neglected, curves of the Binder ratio versus the tuning parameter for different size cross at the critical point. In first-order phase transition, negative values for the Binder ratio will be developed~\cite{Binder1984}.

We show the the data of Binder ratio in Fig.~\ref{fig:fig2}. Without the coupling to the Dirac fermions, i.e., $\lambda=0$, Fig.~\ref{fig:fig2}(a) shows Binder ratio for $\Gamma_x=4.4$. The curves of Binder ratio belonging to various system sizes cross at a point.   This demonstrates that a continuous phase transition (of a $(2+1)$ Ising universality) occurs for this set of parameters. Fig.~\ref{fig:fig2}(b) is also without coupling to the Dirace fermions, it shows the curves of Binder ratio for $\Gamma_x=4.0$, different system sizes develop small negative values, signifying that the parameter set is close to the CTP of the pure boson model. This is consistent with the previous literature~\cite{Kato2015}. After introducing the coupling to the Dirac fermions with $\lambda=0.3$ but still keeping $\Gamma_x=4.0$, Fig.~\ref{fig:fig2}(c) shows that obvious negative values appear in the Binder ratio, indicating the appearance of the first-order phase transition. Moreover, as $L$ increases, the valuea of the Binder ratio tend to diverge, which is a typical signature of the first order transition. Such results reveal that the boson continuous phase transition is changed into a first-order phase one by coupling to the Dirac fermions.

\begin{figure}[htp!]
	\includegraphics[width=\columnwidth]{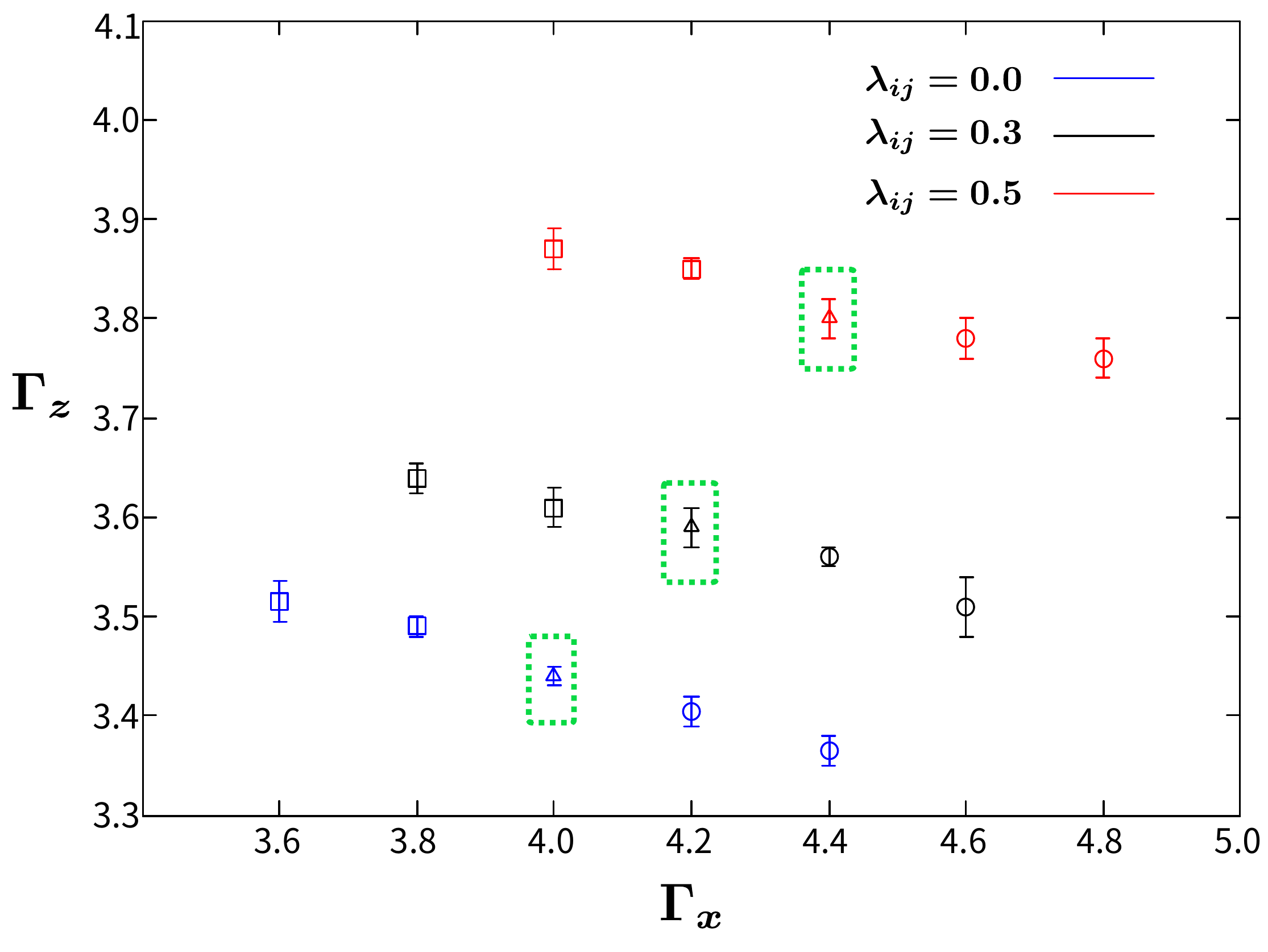}
    \caption{Phase diagram obtained by DQMC near tricritical points in the presence of different coupling $\lambda=0.0(\text{blue})$, $0.3(\text{black})$ and $0.5(\text{red})$. $\Box$ represents first order transition , $\bigtriangleup$ with green border represents the CTPs and $\circ$ represents continue phase transition.}
	\label{fig:fig3}

\end{figure}

To further illustrate this result, we scan the $\Gamma_x$--$\Gamma_z$ phase diagram with different values of the fermion-boson coupling $\lambda$, the phase diagram of pure boson model $(\lambda=0)$ is repeated and consistent with the result in Ref.~\onlinecite{Kato2015}. The phase boundaries are obtained by inspecting the behavior of Binder ratio as shown in Fig.~\ref{fig:fig2}. In Fig.~\ref{fig:fig3}, the blue point is the phase boundary for the pure spin model, and its tricritical point is denoted by a blue triangle; while the black and red points are the phase boundaries in the presence of the coupling to the Dirac fermions, with the coupling strength being $\lambda=0.3$ and $\lambda=0.5$, respectively. One finds that for fixed $\Gamma_x$, the value of $\Gamma_z$ at the phase boundary increases as the coupling strength increases. Moreover, one finds that the regions for the first-order phase transition are extended as $\lambda$ increases. Thus, it seems that the first-order phase transition is apparently enhanced by the fluctuation from the Dirac fermions.

\subsection{Modified mean-field theory for finite-size systems}
\label{sec:iiiB}

To understand the fermion-enhanced first order phase transition, we here develop a modified mean-field theory. In this theory, we focus on the effective potential of the boson field after integrating out the fermion fluctuations. The fermion fluctuations can be truncated from lower bound, since the momentum cannot be smaller than $1/L$ in the lattice model.

We begin with the pure spin (boson) model, whose Hamiltonian is shown in Eq.~(\ref{eq:eq1}) and its mean-field analyses has been reported in Ref.~\onlinecite{Kato2015}. The expectation value of $\sigma_z$ can be decomposed as
\begin{equation}\label{eq:eq7}
\langle \sigma^{z}_{i} \rangle=
\left\{
	\begin{aligned}
		s+\phi_{b} &   &  (i \in A)\\
		s-\phi_{b} &   &  (i \in B)
	\end{aligned}
\right.
\end{equation}
in which $\phi_b$ is the boson dynamical field, $s$ is the background field for $J_a=J_b$ which is just the condition employed in the present work, and $A$, $B$ represents the indices for sublattice. Then by doing the following replacement
\begin{equation}\label{eq:eq8}
\sigma^{z}_{i}\sigma^{z}_{j} \rightarrow \sigma^{z}_{i}\langle \sigma^{z}_{j} \rangle + \langle \sigma^{z}_{i} \rangle \sigma^{z}_{j} - \langle \sigma^{z}_{i} \rangle \langle \sigma^{z}_{j} \rangle,
\end{equation}
one obtains the mean field bosonic Hamiltonian as~\cite{Kato2015}
\begin{eqnarray}\label{eq:eq9}
	\begin{split}	
		\frac{H^{\text{MF}}_{\text{Boson}}}{N} &= (J_{-}s - J_{+}\phi_{b} - 		\Gamma_{z})\sigma^{z}_{A} - \Gamma_{x}\sigma^{x}_{A}  \\
	& +(J_{-}s + J_{+}\phi_{b} -\Gamma_{z})\sigma^{x}_{B} - \Gamma_{x}\sigma^{x}_{B} \\
	& -J_{-}s^{2} + J_{+}\phi_{b}^{2}
	\end{split} 	
\end{eqnarray}
where $J_{\pm}=4(J_a \pm J_b)$. With $J_{-} = 0$ or $J_a = J_b$, this Hamiltonian gives the bosonic free energy per unit cell according to $f_{b}\equiv-\frac{1}{\beta N}\text{logTr}(\text{e}^{-\beta H^{\text{MF}}_{\text{Boson}}})$. At $T=0$, near the phase transition, the free energy can be expanded as function of $\phi_{b}$ as follows
\begin{equation}\label{eq:eq10}
f_{\text{b}} = f_{0}+ \frac{r}{2}\phi_{b}^2 + \frac{u}{4}\phi_{b}^4 + \frac{v}{6}\phi_{b}^6+...,
\end{equation}
in which the coefficient $f_{0}$, $r$, $u$, and $v$ read
\begin{eqnarray}\label{eq:eq11}
\begin{split}
&f_{0} = - \Delta, \\
&\Delta = \sqrt{\Gamma_{x}^2 + \Gamma_{z}^2}, \\
&r = \frac{1}{2}J_{+}(1 - \frac{\Gamma_{x}^{2} J_{+}}{\Delta^3}), \\
&u = \frac{(\Gamma_{x}^2 - 4\Gamma_{z}^2)\Gamma_{x}^{2}J_{+}^{4}}{8\Delta^7}\\
&v = \frac{(12\Gamma_{x}^{2}\Gamma_{z}^{2}-8\Gamma_{z}^{4}-\Gamma_{x}^{4})\Gamma_{x}^{2} J_{+}^{6}}{16\Delta^{11}}
\end{split}
\end{eqnarray}
and the ellipsis represents the higher order terms which are irrelevant and can be neglected. For $\Gamma_x>2\Gamma_z$, $u>0$ and the system hosts a continuous phase transition; while for $\Gamma_x<2\Gamma_z$, $u<0$ and the system hosts the Landau-Devonshire first order phase transition~\cite{Devonshire}. In the continuous case with $u>0$, the order parameter $\phi_b$ develops as $\phi_b=\sqrt{-r/u}$ continuously when $r$ decreases from its critical point $r=0$. In contrast, when $u<0$, $\phi_b$ jumps from zero to $\phi_b=\pm\sqrt{-3u/4v}$ at the transition point $r_t=3u^2/16v$. In addition, when $r<0$, the ordered phase is the only stable phase; when $r>u^2/8v$, the disordered phase is the only stable phase. In between when $0<r<u^2/8v$, both phases can coexist. When $r_t<r<u^2/8v$, the disordered phase is more stable, and when $0<r<r_t$, the ordered phase is more stable.

To explore the effects induced by the coupling to the Dirac fermion, in principle one should consider the boson and fermion fluctuations simultaneously and investigate the renormalization flow on all relevant and marginal operators. However, for the finite-size system, such procedure is quite complex to implement. We have to take a step back and study the influence of the Yukawa coupling on the boson free energy.  The mean-field free energy density reads $f\equiv-\frac{1}{\beta N}\text{logTr}(\text{e}^{-\beta (H^{\text{MF}}_{\text{Boson}}+H_{\text{Fermion}}+H^{\text{MF}}_{\text{Coupling}})})$, in which $H^{\text{MF}}_{\text{Coupling}}$ is the mean-field version of the coupling Hamiltonian~(\ref{eq:eq4}) with $\sigma_z$ being approximated by its mean-field expectation value Eq.~(\ref{eq:eq7}). Note that in $f$, $H_{\text{Fermion}}$ keeps intact as that in Eq.~(\ref{eq:eq2}) and contributions from both valleys are included. At zero temperature, we have
\begin{equation} \label{eq:freeener}
f=f_b+(-)\frac{1}{\pi}\int d^2k \sqrt{\lambda^2 \phi_b^2+2t^2 k^2},
\end{equation}
in which the last term comes from the coupling with the Dirac fermions. In the thermodynamic limit, the range of integral $\int$ is from zero to $\Lambda$ ($\Lambda$ is the ultraviolet cutoff). By explicitly integrating out Eq.~(\ref{eq:freeener}), one finds that the Yukawa coupling between the Dirac fermion and the boson fluctuations can not only change the coefficients in Eq.~(\ref{eq:eq11}), but also generate an additional nonanalytic term $\frac{\lambda^3 |\phi_b|^3}{6 t^2}$. This cubic term is traced back to the gapless Dirac points in the thermodynamic limit.  Actually, at these singular point, the fermion functional integral is ill-defined. 

However, in finite-size systems, fluctuations are truncated from the IR limit by the system size $L$. In this case, a fermion gap appears proportional to $1/L$. Subsequently, after integrating out the fermion fluctuating modes with length scale from $1/L$ to $\Lambda$, the nonanalytic term vanishes and the effective quadratic and quartic coupling in the boson part free energy reads
\begin{eqnarray}\label{eq:eq12}
\begin{split}
	r^{\prime}&=r - \frac{\lambda^{2}(L\Lambda-1)}{4\sqrt{2}\pi L t}, \\
	u^{\prime}&=u + \frac{\lambda^{4}(L\Lambda-1)}{64\sqrt{2}t^{3}\Lambda\pi},
\end{split}
\end{eqnarray}
respectively, with the fermion hopping $t=1$.

\begin{figure}[htp!]
	\includegraphics[width=\columnwidth]{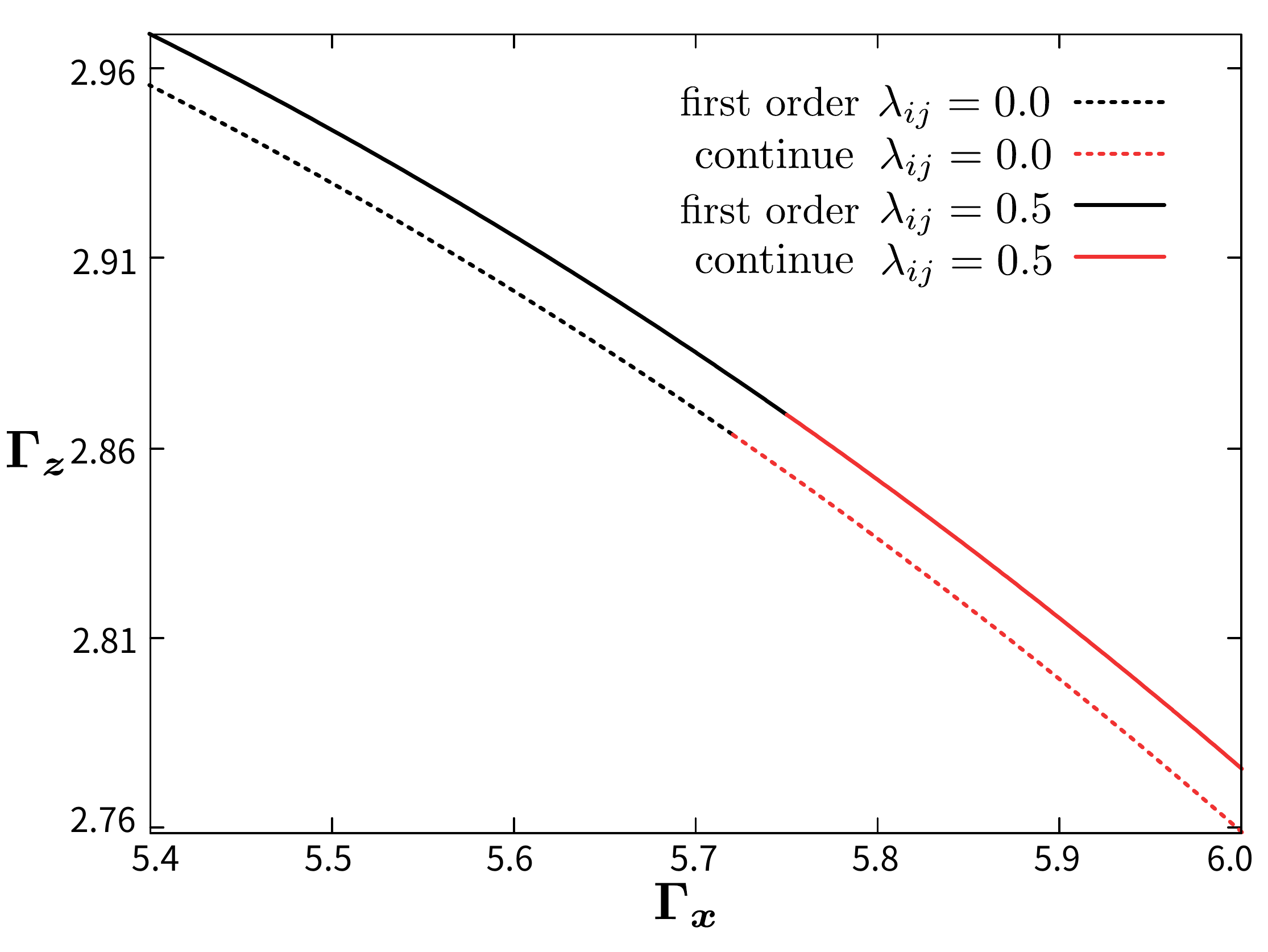}
	\caption{Phase diagram obtained from modified mean-field theory for finite-size systems according to Eq.~(\ref{eq:eq12}) with fixing the coefficient $L=10$, $\Lambda=1$. Black (red) lines represent first order (continue) phase transiton, dot (solid) lines are phase boundaries with the coupling of $\lambda=0.0$ ($0.5$), respectively. }
	\label{fig:fig4}
\end{figure}

At first, we study the change of the phase boundary after turning on the coupling between the Dirac fermions and the dynamical boson field. The phase transition occurs at $r^{\prime}=0$. By substituting this condition and Eq.~(\ref{eq:eq11}) into Eq.~(\ref{eq:eq12}), one finds that for fixed $\Gamma_x$ and $\lambda$, at the phase transition, the value of $\Gamma_z$ changes as
\begin{equation}\label{eq:eq13}
\delta \Gamma_z=\frac{\lambda^{2}(L\Lambda-1) \Delta^{5}}{4\sqrt{2}\pi L t J_{+}^2\Gamma_{x}^2 \Gamma_{z}},
\end{equation}
in which $L\Lambda>1$ since the lattice constant is chosen to be $1$ and other parameters are all positive. Thus, one finds that the value of $\Gamma_z$ increases with $\lambda$ growing. The result is shown in Fig.~\ref{fig:fig4}. It is interesting to see that this result is qualitatively consistent with the DQMC numerical results shown in Fig.~\ref{fig:fig3}.

\begin{figure}[htp!]
	\includegraphics[width=\columnwidth]{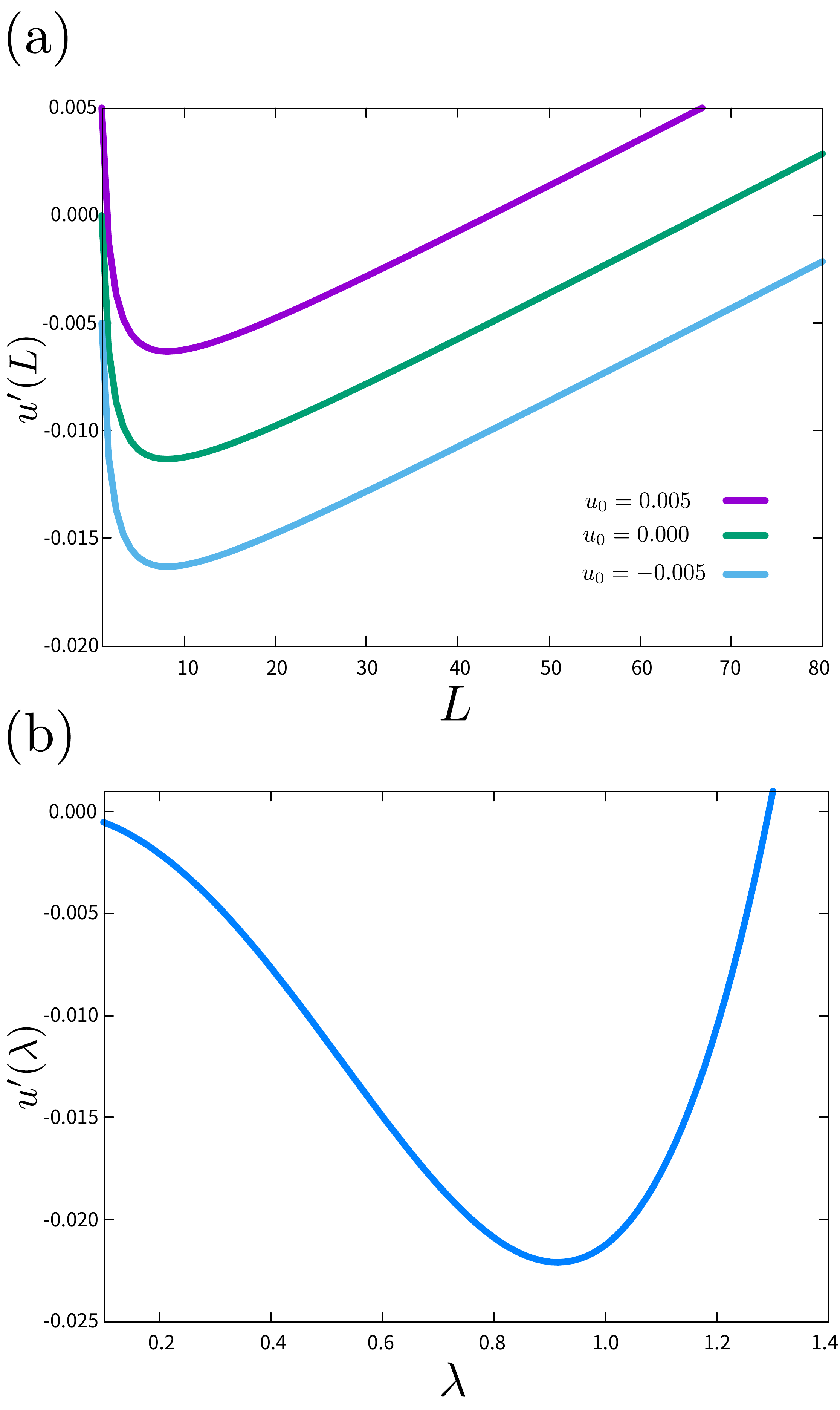}
	\caption{Quartic term $u^{\prime}$ depended on $\lambda$ and $L$ according to Eq.~(\ref{eq:eq14}). (a) $u^{\prime}$ vs $L$ for three types of bosonic quartic contribution $u$ by fixing $\lambda=0.5$.  (b) $u^{\prime}$ vs $\lambda$ by fixing $L=10$.}
	\label{fig:fig5}
\end{figure}

Then, we explore the phase transition properties via this modified mean-field approach. By substituting Eq.~(\ref{eq:eq11}) into Eq.~(\ref{eq:eq12}) and setting $r^{\prime}=0$, one gets
\begin{equation}\label{eq:eq14}
u^{\prime} = u + \left(\Lambda-\frac{1}{L}\right)\left[\frac{L\lambda^4}{64\sqrt{2}t^{3}\Lambda\pi}-\frac{5J^{2}_{+}(3\Gamma^{2}_x-4\Gamma^{2}_z)\lambda^{2}}{48\sqrt{2}\pi t \Delta^{4}}\right].
\end{equation}
According to Eq.~(\ref{eq:eq14}), Fig.~\ref{fig:fig5} explicitly shows the dependence of $u^{\prime}$ on $L$ and $\lambda$. From Fig.~\ref{fig:fig5}(a) one finds that for small system size, $L \sim 10$, $u$ decreases as $L$ increases. This explains the enhancement of the first order phase transition with the increasement of the system size, as shown in Fig.~\ref{fig:fig2} (c). In addition, Fig.~\ref{fig:fig5}(b) shows that for small system sizes, $u$ decreases as $\lambda$ increases. This is consistent with the numerical result that the first order phase transition is enhanced for larger Yukawa coupling, as shown in Fig.~\ref{fig:fig3}. Therefore, as for the model investigated in the paper, the fermion enhanced first-order phase transition is revealed numerically and understood analytically.

\begin{figure}[htp!]
	\includegraphics[width=0.8\columnwidth]{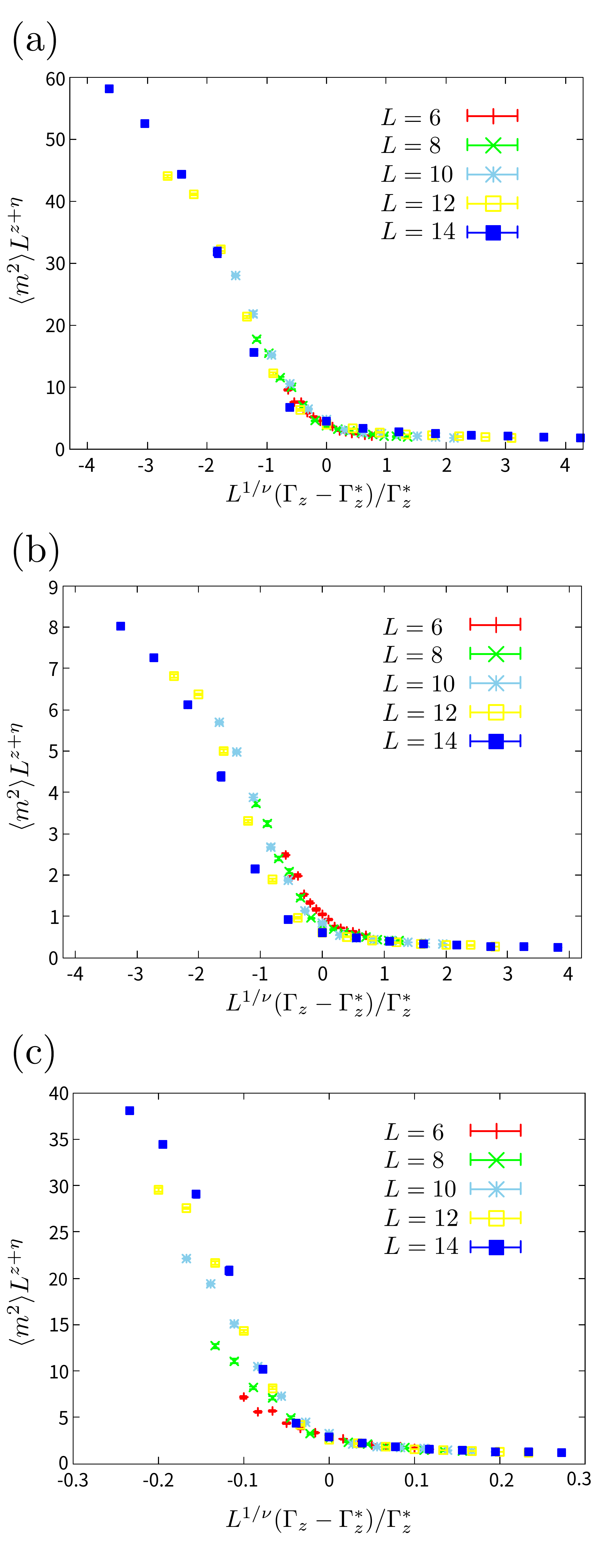}
	\caption{Data collapse of $\langle m^{2} \rangle$ for different critical exponents close to the CTP at $(\Gamma^{*}_x=4.2, \Gamma^{*}_z=3.6)$ with the fermion-boson coupling stregnth $\lambda=0.3$ .(a) Chiral Ising Gross-Neveu CTP universality class, (b) Mean-field Ising tricritical universality class, (c) Chiral Ising Gross-Neveu universality class. The critical exponents used are shown in Tab.~\ref{tab:tab1}.}
	\label{fig:fig6}
\end{figure}
We shall also stress that although this mean-field approach explains the enhancement of the first order phase transition for small system sizes and small Yukawa coupling, there are also limitations in such analysis and open questions remain to be solved. For instance, for any ultraviolet value of $u$, $u^{\prime}$ will change back to positive values. This is contrary to the theoretical prediction that there exists a tricritical point for finite Yukawa coupling. Such limitation can be traced back to the procedure that we do not treat the boson and fermion fluctuations on the same footing, so the ultimate fate of $u'$ in the renomalization flow is still largely unkown. Moreover, since the computational complexity of DQMC scales with a high-power with respect to $L$, numerical calculations for even larger system sizes are increasingly difficult to carry out. For larger $\lambda$, numerical results show apparent unstable results. The reason may be the effects induced by the higher order terms. Therefore, despite that the boson quartic coupling change to positive for large $L$ and $\lambda$ according to the modified mean-field method presented here, it is still a open question to explore the entire parameter region of the fermion enhanced first-order phase transition.



\section{Chiral Gross-Neveu tricritical point}
\label{sec:IV}

\begingroup
\setlength{\tabcolsep}{6pt} 
\renewcommand{\arraystretch}{1.5} 
\begin{table*}[htp!]
	\centering
	\begin{tabular}{c c c c c}
		\hline\hline
		Universality class & $\nu$ & $\eta_{\phi}$ & $\eta_{\psi}$ & $\omega$ \\
		\hline
		Chiral Ising Gross-Neveu CTP in this work & 0.49 & 0.75 & &  \\
		Ising Tricritical point
		
		 (mean-field)~\cite{Kato2015}&1/2 &0 & & \\
		Chiral Ising Gross-Neveu~\cite{Liu2019}& 1.0 & 0.59 & 0.05 & 0.8    \\
		Chiral Ising Gross-Neveu CTP from functional renomalization group\cite{Yin2018} & 0.435 & 0.736 & 0.036 &  \\
		\hline\hline
	\end{tabular}
	\caption{Different critical exponents for the Tricritical point.}
	\label{tab:tab1}
\end{table*}
\endgroup

The fermion fluctuations not only change the type of phase transition but also influence the critical properties. A prevalent example is the chiral universality class, in which the Dirac fermions drive the Wilson-Fisher fixed point for the pure boson model into the Gross-Neveu fixed point\cite{TCLang2013,Mihaila2017, Zerf2017,Gracey2018}. Persistent efforts, including both theoretical and numerical works, have been devoted to unveil the critical properties in these systems~\cite{Herbut2006,Honerkamp2008,Herbut2009,Strack2010,Yao2015,Ihrig2018,Mihaila2017,Zerf2017,YYHe2018,Meng2019,Lang2019,herbutlor,Schuler2019,Liu2019,Gross1974,Rosenstein1993,JanssenHerbut2014,Gracey2018,Knorr2018}. As a generalization of the chiral critical point, the CTP manifest itself when the usual bosonic tricritical point is coupled to the Dirac fermions. Similar to the critical point, it was shown that the fermion fluctuation can drive the bosonic tricritical behavior into a new universality class~\cite{Yin2018}. However, to the best of our knowledge, studies on the the chiral tricritical point (CTP) were hitherto limited in the theoretical approach. Here, we employ the DQMC to explore the critical properties near the chiral tricritical point.

We locate the position of the CTP at $\Gamma^{*}_z$ by the crossing point of Binder ratio which opportunely appear a small negetive value at $\Gamma^{*}_x$, and the obtained CTPs are shown in Fig.~\ref{fig:fig3} as the triangles with green border. There are two relevant directions near the CTP, the one associated with the mass term $r^{\prime}$ and the other the quartic term $u^{\prime}$, both of which are shown in the Eq.~(\ref{eq:eq12}). At the tricritical point, the former dominates.

We compute the order parameter $m^2$ close to the CTP for $\lambda=0.3$ for various lattice sizes. As shown in Fig.~\ref{fig:fig6} (a), by rescaling the curves of $m^2$ versus $\Gamma_z-\Gamma^{*}_{z}$ according to the finite-size scaling form $\langle m^2 \rangle L^{z+\eta}=f(L^{1/\nu}(\Gamma_z-\Gamma^{*}_{z})/\Gamma^{*}_{z})$ with $(\Gamma^{*}_x=4.2,  \Gamma^{*}_z=3.6)$ being the value at the CTP, we find the curves of $m^2$ collapse onto each other when the critical exponents $\nu=0.49$ and $\eta_{\phi}=0.75$ as shown in the first row in Table~\ref{tab:tab1}. As a contrast, we also plot the rescaled $m^2$ curves with the mean-field tricritical exponents $\nu=1/2$ and $\eta=0$ for the pure boson model (second row in Table~\ref{tab:tab1}) which is at the upper critical dimension~\cite{Kato2015} and the chiral Ising critical exponents at the continuous transition (third row in Table~\ref{tab:tab1}) determined from previous DQMC simulation~\cite{Liu2019}, with the corresponding results show in Fig.~\ref{fig:fig6} (b) and (c), respectively. We find that the collapse is obviously better in Fig.~\ref{fig:fig6} (a), and the exponents are close to the predicted chiral Ising Gross-Neveu CTP from functional renormalization group analysis~\cite{Yin2020}, as shown in the fourth row in Table~\ref{tab:tab1} which provide the strong evidence for the existence of CTP. The discrepancy of critical exponent between the first and fourth row in Table~\ref{tab:tab1}  may come from the truncation approximation in the functional renormalization group calculation in the previous literature~\cite{Yin2018} or the finite size scaling in the DQMC simulation in this work.

Comparing the exponents of the CTP with those of the Ising tricritical point as shown in Table.~\ref{tab:tab1}, one finds that a nonzero anomalous dimension of the boson field for the CTP is developed, while it is zero for the Ising tricritical point. The reason is that the $(2+1)$D is the upper critical dimension for the Ising tricritical point, while for the CTP the upper critical dimension cannot be determined from naive power counting of the dimension analysis, since near the Gaussian fixed point, the dimensions of the quartic boson coupling and the Yukawa coupling are different. This behavior may also prohibit the study of the properties of the CTP from usual perturbative renormalization group from the dimension regularization. From this point of view, our present work provides a solid verification of the existence of the CTP and its related critical properties.

\section{Summary}
\label{sec:V}
In summary, we have numerically studied the phase transitions in the Landau-Devonshire model coupled to the Dirac fermions. We find that the interplay of critical fluctuations and finite-size effect can give rise to fermion-enhanced first-order phase transition. This seems in contrary to the theory of the type-II FIQCP. By developing a modified mean-field theory, we show that the reason for this anomalous phenomenon is the interplay between the fermion fluctuations and the finite-size effects, and the fate of the type-II FIQCP for larger system sizes remains to be addressed. Moreover, we have numerically revealed the critical behavior near the chiral Ising Gross-Neveu tricritical point, and for the first time, obtained the new critical exponents therein. Our result demonstrates that the interplay of massless Dirac fermions, critical fluctuations and the finite size effects could trigger a plethora of interesting phenomena and therefore great care is called for when making generalizations. In the future, it will be instructive to explore similar behaviors in other systems with finite Fermi surfaces other than Dirac cones and also interesting to study the full scaling form in these fermion-boson coupled systems, including the other relevant directions.

\section*{Acknowledgement}
YZL and ZYM acknowledge
the support from  the RGC of Hong Kong SAR of China
(Grant Nos. 17303019 and 17301420), MOST through the
National Key Research and Development Program (Grant
No. 2016YFA0300502) and the Strategic Priority Research
Program of the Chinese Academy of Sciences (Grant No.
XDB33000000).  SY is supported by the startup grant (No. 74130-18841229) at
Sun Yat-Sen University. We
thank the Computational Initiative at the Faculty of Science and the Information Technology Services at the University of Hong Kong, and the Tianhe-1A and Tianhe-3 prototype platforms at the National Supercomputer Centers in Tianjin for their technical support and generous allocation of CPU time.

\begin{appendix}

\section{Determinant Monte Carlo Method}
\label{app:appA}

We use determinant quantum Monte Carlo (DQMC) to simulate model which is illustated by Eq.~\eqref{eq:eq4}. We start with the partition function
\begin{equation}\label{eq:eqA1}
	Z = \text{Tr}\{ \text{e}^{-\beta H}\}=\sum_{[\sigma^{z}]}\omega_{B}[\sigma^{z}]\omega_{F}[\sigma^{z}]
\end{equation}
where the configuration space of $[\sigma^{z}]$ is comprised of Ising field. The bosonic part of the partition function is
\begin{eqnarray}\label{eq:eqA2}
\begin{split}
	\omega_{B}=&\text{exp}[-(\Delta\tau J_a \sum_{l}\sum_{\left\langle pq \right\rangle}\sigma^{z}_{p,l}\sigma^{z}_{q,l}- \\
	\Delta\tau J_b \sum_{l}\sum_{\left\langle\left\langle pq \right\rangle\right\rangle}\sigma^{z}_{p,l}\sigma^{z}_{q,l} &-\gamma\sum_{p}\sum_{l}\sigma^{z}_{p,l+1}\sigma^{z}_{p,l}-\Delta\tau\Gamma_z\sum_{p}\sum_{l}\sigma^{z}_{p,l})]
\end{split}
\end{eqnarray}
from $H_{\text{Boson}}$ in Eq.~\eqref{eq:eq1}, when the 2D transvere-field Ising model is mapped to a 3D classical model with $\gamma=-\frac{1}{2}\text{ln}[\text{tanh}(\Delta\tau\Gamma_x)]$.
Meanwile, Fermion part of the partition function is
\begin{equation}\label{eq:eqA3}
	\omega_F=\prod_{\sigma=\uparrow,\downarrow} \det[1+B^{\sigma}_{M}B^{\sigma}_{M-1} \cdots B^{\sigma}_{2}B^{\sigma}_{1}]
\end{equation}

Due to the spin-staggered phase $\text{e}^{i\sigma \phi}$ in $H_{\text{Fermion}}$ term in Eq.(\ref{eq:eq2}), the spin-up determinant $\det[1+B^{\uparrow}_{M}B^{\uparrow}_{M-1} \cdots B^{\uparrow}_{2}B^{\uparrow}_{1}]$ is complex-conjugate to the spin-down one $\det[1+B^{\downarrow}_{M}B^{\downarrow}_{M-1} \cdots B^{\downarrow}_{2}B^{\downarrow}_{1}]$ which means no-sign problem in the system. The $B^{\sigma}_{l}$ can be decomposed into two parts.
\begin{equation}\label{eq:eqA4}
	B^{\sigma}_{l}=\exp(-\Delta\tau H_{\text{Fermion}})\exp(-\Delta\tau H_{\text{Coupling}})
\end{equation}
For $H_{\text{Coupling}}$ term as Fermion-Boson coupling term in Eq.~\eqref{eq:eq3}, it's seperated into four parts such that in each part all the hopping term commute with each other.
For sampling the configuration, we update one site Ising field which the acceptance ratio is expressed as
\begin{equation}\label{eq:eqA5}
r = \frac{\omega_{B}[\sigma^{'}_z]}{\omega_{B}[\sigma_z]}\frac{\omega_{F}[\sigma^{'}_z]}{\omega_{F}[\sigma_{z}]}
\end{equation}
Because $\sigma^{z}$ only have two possible value, bosonic part acception ratio is
\begin{eqnarray}
\begin{split}
\frac{\omega_{B}[\sigma^{'}_z]}{\omega_{B}[\sigma_z]}=&\text{exp}[2(\Delta\tau J_a \sum_{l}\sum_{\left\langle pq \right\rangle}\sigma^{z}_{p,l}\sigma^{z}_{q,l}\\
&-\Delta\tau J_b \sum_{l}\sum_{\left\langle\left\langle pq \right\rangle\right\rangle}\sigma^{z}_{p,l}\sigma^{z}_{q,l} \\
	&-\gamma\sum_{p}\sum_{l}\sigma^{z}_{p,l+1}\sigma^{z}_{p,l}-\Delta\tau\Gamma_z\sum_{p}\sum_{l}\sigma^{z}_{p,l})]
\end{split}
\end{eqnarray}
assuming $B(\tau,0)=B_{1} \cdots B_{\tau}$ and $B(\beta, \tau)=B_{M} \cdots B_{\tau}$, Fermion part is
\begin{equation}
\frac{\omega_{F}[\sigma^{'}_z]}{\omega_{F}[\sigma_{z}]}=\frac{\det[I+B(\beta,\tau)(1+\Delta)B(\tau,0)]}{\det[I+B(\beta,\tau)B(\tau,0)]}
\end{equation}
where $I$ is unit matrix and
\begin{equation}
\Delta=\exp(-\Delta\tau H_{\text{Coupling}}[\sigma^{\prime}_{z}])\exp(\Delta\tau H_{\text{Coupling}}[\sigma_{z}])-1
\end{equation}
in which $\sigma^{\prime}_{z}$($\sigma_{z}$) is updated(original) Ising spin.
\end{appendix}

\bibliographystyle{apsrev4-1}
\bibliography{Fermion}

\end{document}